\documentclass{mem}
\usepackage{natbib}\usepackage{txfonts}\usepackage{balance}
\usepackage{graphicx}
\usepackage[a4paper,breaklinks,dvipdfm]{hyperref}
\idline{84}{1}
\usepackage{txfonts} 

\begin{document}
\def\teff{$T\rm_{eff }$}
\def\kms{$\mathrm {km s}^{-1}$}
\def\ergs{$\mathrm {erg s}^{-1}$}

\title{
New insights on the distant AGN population
}

   \subtitle{}

\author{
A.\,Del Moro\inst{1} 
\and D.~M.\, Alexander\inst{1}
\and J.~R.\, Mullaney\inst{1}
\and E.\, Daddi\inst{2}
\and F.~E.\, Bauer\inst{3,4}
\and A.\, Pope\inst{5}
}

\offprints{A. Del Moro}

\institute{
Durham University, Department of Physics, South Road, DH1 3LE, Durham, UK 
\email{agnese.del-moro@durham.ac.uk}
\and Laboratoire AIM, CEA/DSM-CNRS-Universit\'{e} Paris Diderot, IRFU/Service d' Astrophysique, B\^{a}t. 709, CEA-Saclay, 
91191 Gif-sur-Yvette Cedex, France 
\and Pontificia Universidad Cat\'{o}lica de Chile, Departamento de Astronom\'{\i}a y Astrof\'{\i}sica, Casilla 306, Santiago 22, Chile
\and Space Science Institute, 4750 Walnut Street, Suite 205, Boulder, Colorado 80301, USA
\and Department of Astronomy University of Massachusetts, LGRT-B618, 710 North Pleasant Street, Amherst, MA 01003, USA
}

\authorrunning{Del Moro}
\titlerunning{IR quasars at z$\approx$2}

\abstract{
Current X-ray surveys have proved to be essential tools in order to identify and study AGNs across cosmic time. However, there is evidence that the most heavily obscured AGNs are largely missing even in the deepest surveys. The search for these obscured AGNs is one of the most outstanding issues of extragalactic astronomy, since they are expected to make a major contribution to the high energy peak of the X-ray background (XRB) and might constitute a particularly active and dusty phase of black hole and galaxy evolution. Using a newly developed SED fitting technique to decompose the AGN from the star-formation emission in the infrared band we identify a sample of 17 IR bright quasars at $z=1-3$. For the majority of these sources the X-ray spectrum is well characterised by an absorbed power law model, revealing that $\approx25$\% of the sources are unabsorbed; the remainder of these IR quasars have moderate-to-high absorption ($N_{\rm H}>10^{23}$ cm$^{-2}$), and $\approx$25\% are not detected in the X-rays and are likely to be Compton thick ($N_{\rm H}>10^{24}$ cm$^{-2}$). We therefore find a much higher fraction of obscured than unobscured quasars, indicating that a large fraction of the luminous black hole accretion was very heavily obscured at $z\approx2$.

}
\maketitle{}

\section{Introduction}
The search for obscured and unobscured AGNs has been one of the main focusses of extragalactic astrophysics of the past decades. Since massive black holes (BHs) are hosted in almost 100\% of the galaxy bulges in the local Universe (e.g. \citealt{magorrian1998}), it is clear that tracing the whole AGN population at all redshifts is essential to understand the growth and evolution of galaxies. 

The X-ray band offers a very effective way to trace the growth of BHs, since emission from galaxies is much fainter at these high energies and the radiation can penetrate large amounts of gas and dust without significant absorption. However, even the deepest X-ray surveys fail to observe and identify a large part of the most obscured AGNs (Compton thick, with column densities $N_{\rm H}>10^{24}$ cm$^{-2}$; e.g. \citealt{worsley2005}). A large population of the most obscured AGNs is predicted to reproduce the spectrum of the X-ray background (XRB; e.g., \citealt{comastri1995,gilli2007,treister2009}) at high energies ($E\approx30$ keV) and might represent a particular phase in the BH-galaxy evolutionary scenarios (e.g., \citealt{dimatteo2005,hopkins2006a}).
\begin{figure*}[t!]
\resizebox{\hsize}{!}{
\includegraphics[scale=0.4]{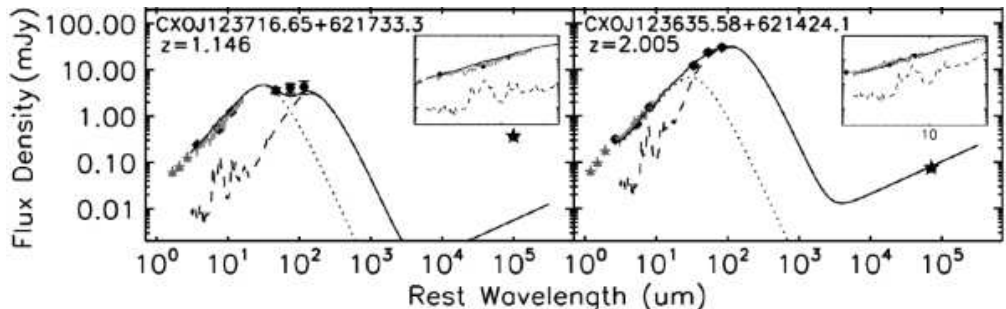}
}
\vspace{-0.8cm}
\caption{\footnotesize Examples of IR SEDs for two of our IR quasars. The AGN component (dotted curve), the star-formation component (dashed curve) and the total best-fitting SED (solid black curve) are shown. The panel on the right of each plot is a zoom into the {\it Spitzer}-IRS spectra ($\lambda\approx5-25\ \mu$m) available for these sources (grey curve). Although the {\it Spitzer}-IRS spectra are not used in the fit to constrain the SEDs, the best-fitting models obtained using only {\it Spitzer} and {\it Herschel} photometric data points agree very well with the IRS spectra \citep{delmoro2013}.}
\label{fig.seds}
\end{figure*}

To identify the most obscured AGNs we thus resort to the infrared (IR) and radio bands (see \citealt{delmoro2013}, for details on the radio AGN selection), where extinction has only small effect on the heated dust emission. However, star formation can dominate the spectral energy distribution (SED) in these bands, and mask the emission from the AGN. Through our newly developed SED decomposition technique (\citealt{delmoro2013}) we are able to disentangle the AGN and star-formation components in the IR band in distant star-forming galaxies ($z\le3$) and effectively identify AGNs that are often undetected in X-rays. We perform this analysis in the GOODS-Herschel North field, revealing a population of luminous IR quasars at $z\approx2$. About 25\% of these bright quasars are not detected in the X-rays and therefore might be very heavily obscured. 

\section{Infrared quasars in $z\approx2$ star-forming galaxies}
\subsection{IR spectral energy distribution}

In our study we use some of the deepest IR {\it Spitzer} and {\it Herschel} data available to date, covering $\approx$160 arcmin$^2$ of the GOODS-North field. We perform a detailed SED decomposition analysis in the IR band for all of the {\it Spitzer} 24 $\mu$m detected sources in the GOODS-Herschel North field with spectroscopic or photometric redshift out to $z=3$ (1825 sources; see \citealt{delmoro2013}, for details). The SEDs are constrained using the {\it Spitzer} 8, 16 and 24 $\mu$m and the {\it Herschel} 100, 160 and 250 $\mu$m data points. We use an empirically defined template to model the AGN IR emission and 5 templates to represent the star-formation emission (Fig. \ref{fig.seds}); the details of the decomposition method and the models used in this analysis are described in \citet{delmoro2013}.

\begin{figure}[t!]
\resizebox{\hsize}{!}{
\includegraphics{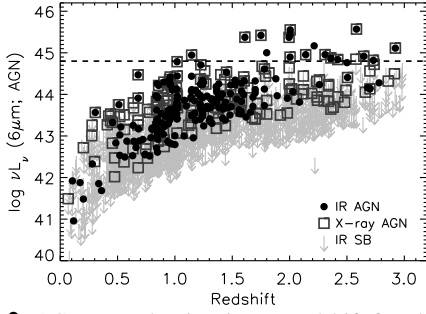}}
\vspace{-0.8cm}
\caption{\footnotesize AGN 6 $\mu$m luminosity vs. redshift for all of the 24 $\mu$m detected sources in the GOODS-Herschel North field. The sources where an AGN component was reliably identified from the SED fitting analysis are represented by black circles, the X-ray detected sources are represented by open squares. Sources where there is no indication of AGN activity in the IR nor in the X-ray band are represented by grey arrows. The dashed line at $\rm log\ L_{6\ \mu m}=44.8$ erg~s$^{-1}$ indicates our selection of IR quasars.}
\label{fig.l6}
\end{figure}

From the resulting best-fitting models we calculate the rest-frame 6 $\mu$m luminosity of the AGN (removing the galaxy contribution), which gives us an estimate of the intrinsic power of the AGN, since the IR emission is only lightly affected by extinction. Where the AGN component is not reliably constrained in our SED fitting, we estimate an upper limit for the AGN luminosity from the total 6 $\mu$m luminosity obtained from the best-fit models (see Fig. \ref{fig.l6}). We significantly identified AGN emission in the IR band in $\approx$12\% (212 sources) of the 24 $\mu$m detected sources (Del Moro et al. 2013, in prep.).
\subsection{IR quasars identification}

From the results obtained from our SED decomposition we identified a sample of 17 bright AGNs with mid-IR luminosity $\rm log\ L_{6\ \mu m}\ge 44.8$ erg~s$^{-1}$ at redshift $z=1-3$ (see Fig. \ref{fig.l6}), corresponding to an intrinsic luminosity in the X-ray band of $\rm L_{X}>10^{44}$ erg~s$^{-1}$ (using the $\rm L_X-L_{6\ \mu m}$ relation for local unabsorbed AGN, e.g., from \citealt{lutz2004}), i.e. in the quasar regime. Almost all of these sources (14/17; i.e. $\approx$82\%) have spectroscopic redshift identification from optical and near-IR spectroscopy and 5 of them are classified as broad-line (BL) AGN in the optical band. For $\approx$53\% of our IR quasars (9/17) {\it Spitzer}-IRS spectra are available (e.g. \citealt{kirkpatrick2012}), which agree remarkably well with our best-fitting models for the source SEDs (see Fig. \ref{fig.seds}); this is an important confirmation that our SED fitting approach is reliable and that the mid-IR AGN luminosities measured from the AGN components are fairly accurate.   

\section{Heavily obscured accretion at $z\approx2$}

Most of the IR quasars in our sample (13/17; $\approx$76\%) are detected in the deep 2 Ms {\it Chandra} data (\citealt{alexander2003}), while 4 IR quasars remain undetected in the X-ray band. 
On the basis of the X-ray spectral analysis we have 10 X-ray luminous AGNs ($\rm L_X > 10^{43}$ erg~s$^{-1}$), which have well characterised X-ray spectral properties ($\Gamma\approx 1.8$ and $\rm N_H < 2\times 10^{23}$ cm$^{-2}$); all of these sources have $>$400 X-ray counts (including the 5 quasars classified as BL AGNs). These sources comprise unabsorbed and moderately absorbed quasars and are easily detected in moderate-depth X-ray surveys (but require deep X-ray surveys to accurately characterise the moderately absorbed systems). 
The remaining 3 X-ray detected quasars have $\rm L_X < 10^{43}$ erg~s$^{-1}$ and flatter X-ray spectral slopes ($\Gamma<1$) than those observed for the more luminous AGNs; the X-ray spectral properties of these sources are more difficult to measure, most likely because (1) they are fainter and have poorer count statistics than the X-ray luminous quasars (all have $<$400 X-ray counts) and (2) they are probably more heavily obscured and cannot be easily characterised by a simple absorbed power law model, since components such as reflection will affect the shape of the X-ray emission (see \citealt{alexander2011}, for a likely similar source population). The 4 X-ray undetected IR bright quasars are only identified here as luminous AGNs on the basis of the presence of AGN emission in the infrared SED. 

\begin{figure}[t!]
\includegraphics[scale=0.3]{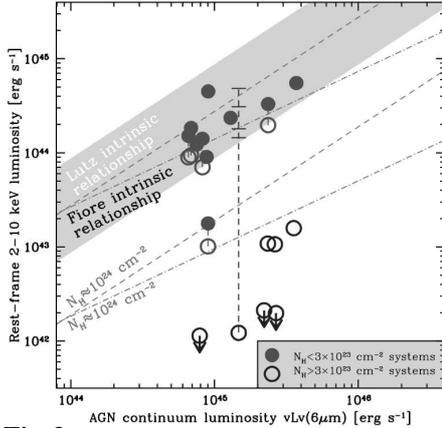}
\vspace{-0.3cm}
\caption{\footnotesize Rest-frame 6 $\mu$m luminosity vs. X-ray luminosity for our IR quasars. Sources with well characterised X-ray spectra and moderate absorption ($\rm N_H<3\times10^{23}$ cm$^{-2}$) are represented by filled circles and follow the typical $\rm L_X-L_{6\ \mu m}$ relation of local unabsorbed AGNs (shaded region; e.g. \citealt{lutz2004,fiore2009}). X-ray undetected sources, as well as the flat-spectrum X-ray sources, are represented by black open circles; the X-ray luminosity and upper limits measured for these sources put them in the Compton-thick region of the $\rm L_X-L_{6\ \mu m}$ plot, which indicates that these sources are likely to be very heavily obscured.
}
\label{fig.l6lx}
\end{figure}
In Figure \ref{fig.l6lx} we compare the X-ray luminosity and upper limits measured for our IR quasars with the mid-IR luminosity $\rm L_{6\ \mu m}$ obtained from the IR SED fitting, and find that the moderately obscured ($\rm N_H<2\times10^{23}$ cm$^{-2}$) and unobscured sources follow nicely the $\rm L_X-L_{6\ \mu m}$ relation observed for local unabsorbed AGNs (e.g. \citealt{lutz2004,fiore2009}). The remaining quasars lie far below the relation, indicating that the X-ray luminosity of these sources must be highly suppressed, possibly by Compton-thick material along the line of sight.

\section{Discussion}

Amongst our sample we find that 5 of the 17 IR quasars are identified as BL AGNs; these sources are similar to those typically found in optical quasar searches such as SDSS. The other 13 sources are obscured quasars, of which 5 are moderately obscured and 7 are likely to be heavily obscured. This means that there are potentially $\approx$2.5 times more obscured quasars than unobscured quasars in our sample. This fraction is much higher than those found in many previous studies ($<$1; e.g. \citealt{ueda2003, lafranca2005, hasinger2008}), indicating that most of the luminous BH growth was very heavily obscured at $z\approx1-3$. 

\section{Summary}

Through detailed SED analysis in the IR band of all of the 24 $\mu$m detected sources in the GOODS-Herschel North field, we identify a population of IR bright quasars at $z\approx2$. The X-ray spectral analyses reveal that moderate-to-heavily obscured AGNs outnumber unabsorbed AGNs by a factor of $\approx$2.5, a much higher fraction than those found in previous studies, suggesting that the IR selection of bright quasars is able to reveal the most obscured accretion at $z\approx2$. 
We are now extending our analyses to the GOODS-South field to increase the source statistics and the constraints on the space density and properties of these highly obscured IR quasars.

\begin{acknowledgements}
We gratefully acknowledge support from the STFC Rolling Grant (ADM; DMA). 
This work is based on observations made with Herschel, a ESA 
Cornerstone Mission with significant participation by NASA.
\end{acknowledgements}




\end{document}